\documentclass[aps,prb,twocolumn,reprint,10pt,amssymb,showpacs,unsortedaddress,superscriptaddress,floatfix]{revtex4-1}
\usepackage{dcolumn}
\usepackage{graphicx}
\usepackage{bm}
\usepackage{textcomp}
\usepackage{amsmath}
\usepackage{natbib}

\begin{document}
\title{The Role of a Polariton Bath in the Emission Spectrum of an Open Nanocavity-Quantum Dot System}
\date{\today}
\author{Nicol{\'a}s \surname{Moure}}
\affiliation{Instituto de F{\'i}sica, Universidad de Antioquia, Medell{\'i}n, Colombia}
\author{Herbert \surname{Vinck-Posada}}
\affiliation{Departamento de F{\'i}sica, Universidad Nacional de Colombia, Bogot{\'a}, Colombia}
\author{Boris A. \surname{Rodr{\'i}guez}}
\affiliation{Instituto de F{\'i}sica, Universidad de Antioquia, Medell{\'i}n, Colombia}

\begin{abstract}

We investigate the effect of a polariton bath on the photoluminescence (PL) spectrum in confined nanocavity-quantum dot (nC-QD) systems. We model the nC-QD system as a two-level exciton in strong coupling with a single photonic cavity mode interacting with its environment. The non-hamiltonian processes induced by the environment are taken into account via a Born-Markov master equation which includes: gain and loss of excitons and photons into/out of the cavity, a dephasing mechanism produced by phonon scattering in the semiconductor lattice and gain and loss of polaritons due to the already mentioned polariton bath.  In order to validate our phenomenological model, we fit an experimental spectrum to extract the values of all parameters appearing in the master equation. Our results show that polariton pumping and loss rates are comparable to the other parameters and therefore we have a first evidence that the polariton bath we proposed has a significant role in the dynamics of a nC-QD system.

\end{abstract}

 

\maketitle

\section{Introduction}

Nanocavity quantum electrodynamics (nC-QED)---the complex interplay between the methods of quantum optics and the specific nature of solid state systems---has recently emerged as one of the most active and promising research fields \cite{kavokin2011microcavities,Elena2010microcavity}. A single quantum emitter, a quantum dot (QD), coupled to one confined light mode in a photonic crystal, micropillar, or microdisk cavity is one of these nC-QED systems with more perspectives for device applications and fundamental physics such as: single photon sources \cite{Nature419_594}, quantum information processing \cite{NewJPhys13_055025}, strong coupling (SC) \cite{Nature432_197, Nature432_200, PRL101_083601, JPhysCondensMatter23_265304}, non-classical light \cite{PRL98_117402,NatPhys4_859}, strong optical nonlinearities \cite{Nature450_862,Science320_769}, and control of cavity reflectivity  \cite{Nature450_857}. More recently, it has been reported controllable all-optical switching between laser pulses at the single photon level using a QD embedded in a photonic crystal cavity \cite{PRL108_093604,PRL108_227402}.\par

One of the most remarkable SC phenomena in nC-QED systems is the polariton physics \cite{kavokin2011microcavities, NatPhoton5_275}. A polariton quasiparticle is an entangled state  between semiconductor excitons and cavity mode photons that arise from the strong-coupling regime of the light-matter interaction. Despite more than twenty years of conducting theoretical and experimental research, the community does not agree yet with the main physics behind polariton phenomena \cite{NatPhoton6_2,NatPhoton6_205}. The large ground state occupation with spontaneous coherence build up can be understood in terms of a polariton laser \cite{Bloch1,PRB_gonzalez,us_arXiv1205.2719,NatPhoton6_2} or as a non-equilibrium Bose-Einstein condensate \cite{Dang,Yamamoto,NatPhoton6_205}.\par

Despite that SC is usually identified through their characteristic anticrossing in the photoluminescence emission spectrum \cite{Nature432_197}, detailed theoretical works have shown that this feature is not an unique fingerprint of SC \cite{PRL101_083601}. The line splitting in the spectrum depends mainly on the interplay of dissipative and dephasing rates and therefore when analyzing an experimental spectrum a comprehensive modeling theory is needed.  A surprising example of agreement between theory and experiment was done by Laucht {\it et al.} in Ref. \onlinecite{PRL_finley}. By fitting the theoretical spectrum with the experimental one, they extracted the whole set of parameters that characterize the interaction mechanisms between the nC-QD system and the environment: gain and loss of excitons and photons into/out of the cavity and a dephasing mechanism due to phonon scattering in the semiconductor lattice.\par

Motivated by the polariton laser experimental results \cite{Bloch1}, recent theoretical works have proposed \cite{PRB_gonzalez,CondMatters} that some of the coherence properties of nC-QED systems can be understood in terms of an effective pumping of polaritons. In Ref. \onlinecite{PRB_gonzalez}, the authors have developed a finite system theory of a multilevel QD which qualitatively reproduces the experimental polariton laser threshold. A Jaynes-Cummings approach was used in Ref. \onlinecite{CondMatters} to show how the incoherent polariton pumping is able to both sustain a large number of photons inside the cavity with Poisson-like statistics, and induce a separable exciton--photon state in the stationary limit. In addition, the authors also have shown that the polariton pumping is unable to modify the dynamical regimes of the system. Despite the huge literature on polaritons in nC-QED systems, a deeper discussion of a comprehensive polariton bath theory is required.\par

In this work, we propose a master equation with a polariton bath, in addition to the already reported processes in the literature\cite{PRB_tejedor,PRB_gonzalez,CondMatters,PRL_finley}. Some of the advantages of this procedure are: the recognition of polariton loss as a relevant scattering process, the improvement in the definition of polaritonic transition operators as discussed in section \ref{sebsec:master}, and the possibility to obtain a better agreement when fitting an experimental spectrum. In view of the high degree of complexity encountered in solid state systems, this could represent another step towards a better understanding of the nC-QD system.\par

The rest of the paper is organized as follows. In Sec. \ref{sec:Th-Mod}, we describe the  theoretical model and write down the master equation with all the processes involved. We also outline the Quantum Regression Theorem (QRT) equations, which permit us to calculate the emission spectrum beyond the linear regime. In Sec. \ref{sec:num-re},  numerical results for the luminescence spectrum are shown. We investigate the influence of our definition of the polariton transition operators and discuss the different approximations to the first excitation manifold found in the literature. In addition, the polariton pumping and loss rates by fitting an experimental spectrum are found. Concluding remarks are presented in Sec. \ref{sec:conclusions}. Finally, the appendices give some details about the density operator matrix elements, the QRT equations, and the approximations to the linear regime.\par

\section{Theoretical Model}\label{sec:Th-Mod}
We consider a single quantum dot in interaction with one photonic mode of a semiconductor nC. For the purposes of our work we treat the QD as a two-level system (monoexcitonic regime). This model is customary in the literature and captures the main physics of the system\cite{PRB_tejedor-fermions,PRB_tejedor,PRL_finley,CondMatters}. Hence, an appropriate way of describing the intrinsic dynamics of the nC-QD system is by means of the Jaynes-Cummings hamiltonian\cite{Jaynes_Cummings} 
\begin{equation}\label{Hs}
H_{\mathcal{S}}=\dfrac{\hbar\omega_{x}}{2}\sigma_{z}+\hbar\omega_{c}a^{\dagger}a+\hbar g(\sigma_{+}a+\sigma_{-}a^{\dagger}), 
\end{equation}
where $\sigma_{+}$, $\sigma_{-}$ and $\sigma_{z}$ are pseudo-spin operators for the QD with ground state $|G\rangle$ and excited (exciton) state $|X\rangle$ separated by the exciton energy $\hbar\omega_{x}$. $a^{\dagger}$ ($a$) is the creation (annihilation) operator of the nC mode with frequency $\omega_{c}$, and $g$ is the coupling strength between the photonic mode and the exciton. We are interested in the strong coupling regime. In this regime, the ``dressed'' states of the system
\begin{equation}\label{dressed}
\begin{split}
 |n,+\rangle&= \cos\left(\frac{\phi_{n}}{2}\right)|Xn-1\rangle+\sin\left(\frac{\phi_{n}}{2}\right)|Gn\rangle,\\
 |n,-\rangle&=-\sin{\left(\frac{\phi_{n}}{2}\right)}|Xn-1\rangle+\cos{\left(\frac{\phi_{n}}{2}\right)}|Gn\rangle,\\
 \phi_{n}&=\tan^{-1}(2g\sqrt{n}/\Delta),
\end{split} 
\end{equation}
represent quasi-particle states, where $|Gn\rangle$, $ |Xn-1\rangle$ are the ``bare'' states of the nC-QD system and $\Delta = \omega_x-\omega_c$ is the detuning. These quasi-particles are called polaritons and their existence becomes evident by looking at the level anti-crossing in the emission spectra\cite{PRL_yamamoto}.\\

\subsection{Master equation}\label{sebsec:master}
In addition to loss and gain of excitons and photons, and a dephasing mechanism already considered in the literature \cite{PRL_finley,PRB_tejedor}, we propose a polariton bath that accounts for gain and loss of polaritons into and out of the system. Theses processes are phenomenologically modeled by the system-environment interaction hamiltonian

\begin{equation}
H_{P}^{\mathcal{SE}}=\sum_{i,j,n}\sum\limits_{R}\hbar\,\mu_{ijn}^{R}(b_{ijn,R}^{\dagger}P_{ij}^{(n)}+b_{ijn,R}P_{ij}^{(n)\dagger}),
\end{equation}
where $b_{ijn,R}$, $b_{ijn,R}^{\dagger}$ are bosonic operators for the bath and $P_{ij}^{(n)}$, $P_{ij}^{(n)\dagger}$ are polaritonic transition operators defined by 
\begin{equation}
\begin{split}\label{Pijn}
 P_{ij}^{(n)}&=\sqrt{n}\,|n-1,j\rangle\langle n,i|\qquad\bigl(\Lambda_{n}\rightarrow\Lambda_{n-1}\bigr); \\
 P_{ij}^{(n)\dagger}&=\sqrt{n}\,|n,i\rangle\langle n-1,j|\qquad\bigl(\Lambda_{n-1}\rightarrow\Lambda_{n}\bigr),
 \end{split} 
\end{equation}
where $i,j\in\{+,-\}$, $n=1,2,\dots, \infty$, and $\Lambda_{n}=\{|Gn\rangle,|Xn-1\rangle\}$ is the  $n$th excitation manifold. $H_{P}^{\mathcal{SE}}$ takes into account situations in which a boson from the reservoir is absorbed ($b_{ijn,R}$) causing a polaritonic transition $|n-1,j\rangle\rightarrow|n,i\rangle$ ($P_{ij}^{(n)\dagger}$) along with the opposite processes. Both channels are controlled by a coupling constant $\mu_{ijn}^{R}$. In previous works \cite{CondMatters,PRB_gonzalez}, similar transition operators to those on (\ref{Pijn}) were defined. However, we have introduced the factor $\sqrt{n}$ to account for the non-linear behavior of polariton gain and loss rates with excitation number as can be expected if we look at their photonic analogs (see Eqs. (\ref{matrix})). Additionally, the polariton number operator defined by $N_P |n,\pm\rangle = n |n,\pm\rangle$, can be written as $N_P=(1/2)\sum_{ijn}{P_{ij}^{(n)\dagger} P_{ij}^{(n)}}$, which somehow justifies our election of the $\sqrt{n}$ factor. \par

Following the Born-Markov formalism \cite{Breuer,Schl} and using a system-environment interaction hamiltonian containing $H_{P}^{\mathcal{SE}}$, we arrive to a master equation for the density operator of the nC-QD system:
\begin{widetext}
\begin{align}\label{master}
  \frac{d\rho}{dt}=&-\frac{i}{\hbar}[H_{S},\rho] +\frac{\gamma_{\phi}}{2}(\sigma_{z}\rho\sigma_{z}-\rho)
  +\frac{P_{a}}{2}(2a^{\dagger}\rho a-aa^{\dagger}\rho-\rho aa^{\dagger})
  +\frac{\gamma_{a}}{2}(2a\rho a^{\dagger}-a^{\dagger}a\rho-\rho a^{\dagger}a)\notag\\
  +&\frac{P_{\sigma}}{2}(2\sigma_{+}\rho\sigma_{-}-\sigma_{-}\sigma_{+}\rho-\rho\sigma_{-}\sigma_{+})
  +\frac{\gamma_{\sigma}}{2}(2\sigma_{-}\rho \sigma_{+}-\sigma_{+}\sigma_{-}\rho-\rho\sigma_{+}\sigma_{-}) \\
  +&\frac{P_{p}}{2}\sum_{ijn}\left(2P_{ij}^{(n)\dagger}\rho P_{ij}^{(n)}-P_{ij}^{(n)}P_{ij}^{(n)\dagger}\rho-
  \rho P_{ij}^{(n)}P_{ij}^{(n)\dagger}\right)
  +\frac{\gamma_{p}}{2}\sum_{ijn}\left(2P_{ij}^{(n)}\rho P_{ij}^{(n)\dagger}-P_{ij}^{(n)\dagger}P_{ij}^{(n)}\rho-
  \rho P_{ij}^{(n)\dagger}P_{ij}^{(n)}\right).\notag
\end{align}
\end{widetext}

\begin{figure}[b]
\includegraphics[scale=1.05]{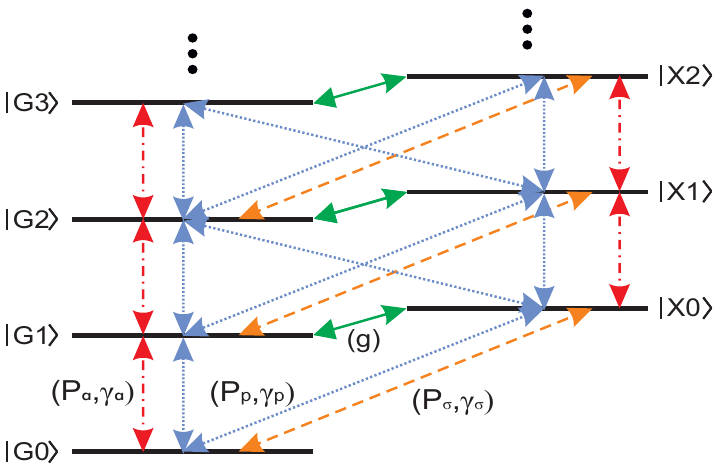}
\caption{\label{fig:levels} (Color online) Transitions between ``bare'' energy levels induced by the hamiltonian and non--hamiltonian processes. Continuous green arrows represent transitions due to the nC-QD interaction. Dot--dashed  red arrows show photon gain and loss transitions. Dashed orange arrows depict transitions caused by the exciton pumping and loss processes. Finally, the dotted blue arrows correspond to the polariton bath induced transitions (these transitions can be better visualized using the ``dressed'' states basis). The dephasing mechanism is not displayed since it does not cause transitions between energy levels, it only modifies the non-diagonal density matrix elements.}
\end{figure}

In this master equation, $\gamma_{\phi}$ is a pure dephasing rate induced by constant scattering processes between the QD and the semiconductor lattice phonons, $P_{a}$ represents the photonic pumping rate caused by the emission of weakly coupled QD's present in the nC, $\gamma_{a}$ is the cavity loss rate (coherent emission), $P_{\sigma}$ is the rate at which excitons are incoherently pumped by optical or electrical excitation, $\gamma_{\sigma}$ is the exciton decay rate (spontaneous emission), and finally $P_{p}$ and $\gamma_{p}$ are respectively, the rates at which polaritons are created and lost (see Fig. \ref{fig:levels}).\par


Eq. (\ref{master}) can be transformed in an infinite set of linear differential equations for the density operator matrix elements on the direct product basis $\{|Gn\rangle,|Xn\rangle\}$. When the system is initially in its ground state, there is an independent set of equations for the nC-QD  populations $ (\rho_{Gn,Gn},\rho_{Xn,Xn})$ and coherences between levels in the same excitation manifold $\Lambda_{n+1}$ ($\rho_{Gn+1,Xn}$). This set of linear equations is found in Appendix \ref{Matrix} (Eqs. \ref{matrix}). It is useful for practical computational purposes and it provides a better understanding of the effect caused by system-environment interaction processes. From Eqs. (\ref{matrix}) it becomes clear that the overall effect of these processes is to cause transitions between excitation manifolds and to reduce coherence in the system.

\subsection{Emission spectrum}
When dealing with experimental samples of the nC-QD system, the emission spectrum is one of the most accessible observables. Assuming that the principal cause of light emission of the system is due to cavity leakage, the spectral function in the stationary state is given by
\begin{equation}\label{S(w)}
S(\omega)\propto\lim_{t\rightarrow\infty}\text{Re}\int_{0}^{\infty}d\tau e^{-(\gamma+i\omega)\tau}\langle a^{\dagger}(t+\tau)a(t)\rangle,
\end{equation}
which is the Fourier transform of the cavity field correlation function $G^{(1)}(t,\tau)=\langle a^{\dagger}(t+\tau)a(t)\rangle$ weighted by the factor $e^{-\gamma\tau}$ to account for additional broadening of the spectral lines due to the finite resolution of the detectors\cite{PRL_finley}.\par

Our calculation of the correlation function closely resembles the one by Tejedor and co-workers \cite{PRB_tejedor}. First, we write operator $a^{\dagger}$ as 
\begin{equation}\label{a}                                                                                                 a^{\dagger}=\sum_{n}\sqrt{n+1}(a^{\dagger}_{Gn}+a^{\dagger}_{Xn}),                                                        \end{equation}
where $a^{\dagger}_{Gn}=|Gn+1\rangle\langle Gn|$ and $a^{\dagger}_{Xn}=|Xn+1\rangle\langle Xn|$.  Using the master equation (Eq. (\ref{master})) it is possible to find an independent set of linear differential equations including the expectation values of $a^{\dagger}_{Gn}$ and $a^{\dagger}_{Xn}$ (see Appendix \ref{QRT}, Eqs. (\ref{QRTeqs})).
According to the Quantum Regression Theorem (QRT), time evolution for expectation values $\langle a_{Gn}^{\dagger}(t+\tau)a(t)\rangle$, $\langle a_{Xn}^{\dagger}(t+\tau)a(t)\rangle$, $\langle \sigma_{n}^{\dagger}(t+\tau)a(t)\rangle$, and $\langle \varsigma_{n}(t+\tau)a(t)\rangle$, is also governed by Eqs. (\ref{QRTeqs}) with initial conditions
\begin{equation}\label{QRTinis}
\begin{split}
 \langle a_{Gn}^{\dagger}(t)a(t)\rangle&=\sqrt{n+1}\,\rho_{Gn+1,Gn+1}(t),\\
 \langle a_{Xn}^{\dagger}(t)a(t)\rangle&=\sqrt{n+1}\,\rho_{Xn+1,Xn+1}(t),\\
 \langle \sigma_{n}^{\dagger}(t)a(t)\rangle&=\sqrt{n+1}\,\rho_{Gn+1,Xn}(t),\\
 \langle \varsigma_{n}(t)a(t)\rangle&=\sqrt{n}\,\rho_{Xn,Gn+1}(t),
\end{split}
\end{equation}
where $t$ stands for the time in which the system has reached the stationary state, and operators $\sigma^{\dagger}_{n}=|Xn\rangle\langle Gn|$ and $\varsigma_{n}=|Gn+1\rangle\langle Xn-1|$ were introduced in order to obtain  a closed set of operators. After solving Eqs. (\ref{QRTeqs}) with initial conditions given in Eqs. (\ref{QRTinis}), we can compute the first-order correlation function as
\begin{equation}\label{g1}
\begin{split}
 G^{(1)}(t,\tau)=\sum\limits_{n}&\sqrt{n+1}\;\left(\langle a_{Gn}^{\dagger}(t+\tau)a(t)\rangle \right.\\
 &\left.+\;\langle a_{Xn}^{\dagger}(t+\tau)a(t)\rangle\right).
\end{split}
\end{equation} 
Now, with the knowledge of $G^{(1)}(t,\tau)$, the emission spectrum is obtained via Eq. (\ref{S(w)}). The main advantage of the previous approach is that we can calculate the emission spectrum beyond the first excitation manifold  without doing any further approximations.

\section{Numerical results}\label{sec:num-re}
\subsection{Polariton pumping and emission spectrum}
In this section, we show the effect in the emission spectrum caused by the excitation manifold dependent factor in the polariton ladder operators definition (Eqs. (\ref{Pijn})). In previous works\cite{CondMatters,PRB_gonzalez}, this factor was not introduced which then leaves the effective polariton pumping and loss rates independent from the excitation number. From Eqs. (\ref{matrix}), it can be seen that effective photonic rates grow with the photon number $n$, a behavior that we reproduce for polariton rates by introducing the factor $\sqrt{n}$ in Eqs. (\ref{Pijn}).\par

\begin{figure}[b]
\includegraphics[scale=0.42]{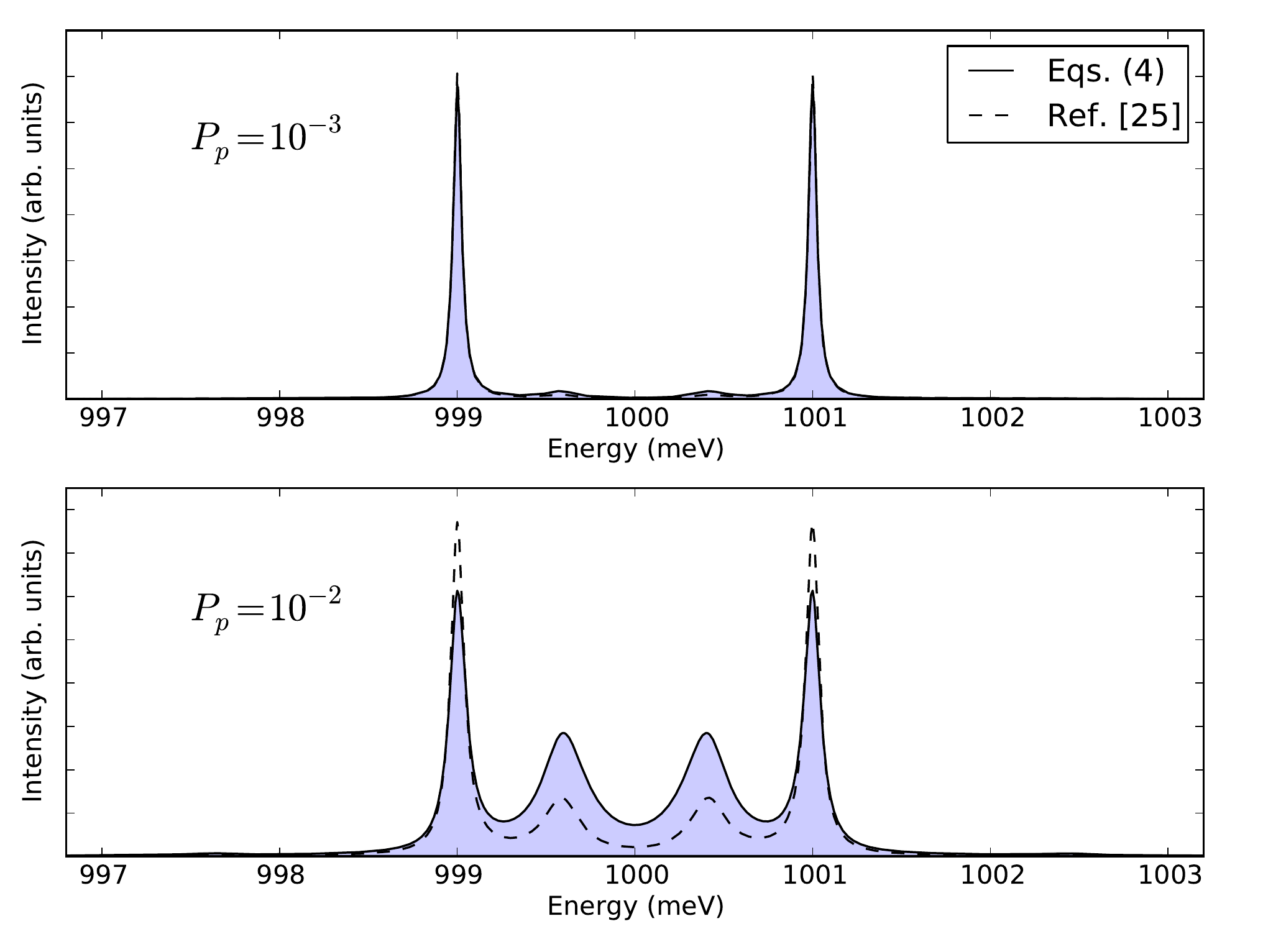}
\caption{\label{fig:2} Emission spectra using two definitions for the polariton ladder operators for $P_{p}=10^{-3}\,$meV$/\hbar$ (top) and $P_{p}=10^{-2}\,$meV$/\hbar$ (bottom). Continuous shadowed lines represent spectral functions calculated with Eqs. (\ref{Pijn}), whereas dashed lines correspond to emission spectra calculated without the factor $\sqrt{n}$ as is used in Ref. \onlinecite{CondMatters}. The other parameters in the master equation were fixed to $\omega_{c}=1000\,$meV$/\hbar$, $g=1\,$meV$/\hbar$, $\gamma_{a}=0$.$1\,$meV$/\hbar$, and the other ones to zero $\Delta=P_{a}=P_{\sigma}=\gamma_{\sigma}=\gamma_{\phi}=0$.}
\end{figure}

In Fig. \ref{fig:2}, we show emission spectra calculated with and without the factor $\sqrt{n}$ in the ladder operators definition. For low polariton rates ($P_{p}=10^{-3}\,$meV$/\hbar$), the difference between both spectral lines is almost negligible. This is expected since for low pumping rates, the population for excitation manifolds higher than $\Lambda_{1}$ are much less than one, making the dependence with $n$ almost irrelevant. However, for higher pumping rates ($P_{p}=10^{-2}\,$meV$/\hbar$), both spectral functions appreciably differ from each other. In fact, the interior peaks have a greater intensity if the spectrum is calculated with ladder operators given by Eqs. (\ref{Pijn}). Again, this is expected since for $n$-dependent polariton pumping, the second excitation manifold populations are higher and the interior peaks correspond to transitions $|2,\pm\rangle\rightarrow|1,\pm\rangle$. Additionally, there is a minor increment in the line widths, since decoherence effects are larger when the factor $\sqrt{n}$ is introduced  (see Eqs. (\ref{matrix})).

\subsection{Approximations to the emission spectrum}\label{sec_apps}
The importance of computing the emission spectrum using the exact dynamics of the first-order correlation function governed by Eqs. (\ref{QRTeqs}) is developed in this section. For this purpose, we fix the polariton pumping and loss rates to zero.\par

\begin{figure}[b]
\includegraphics[scale=0.4]{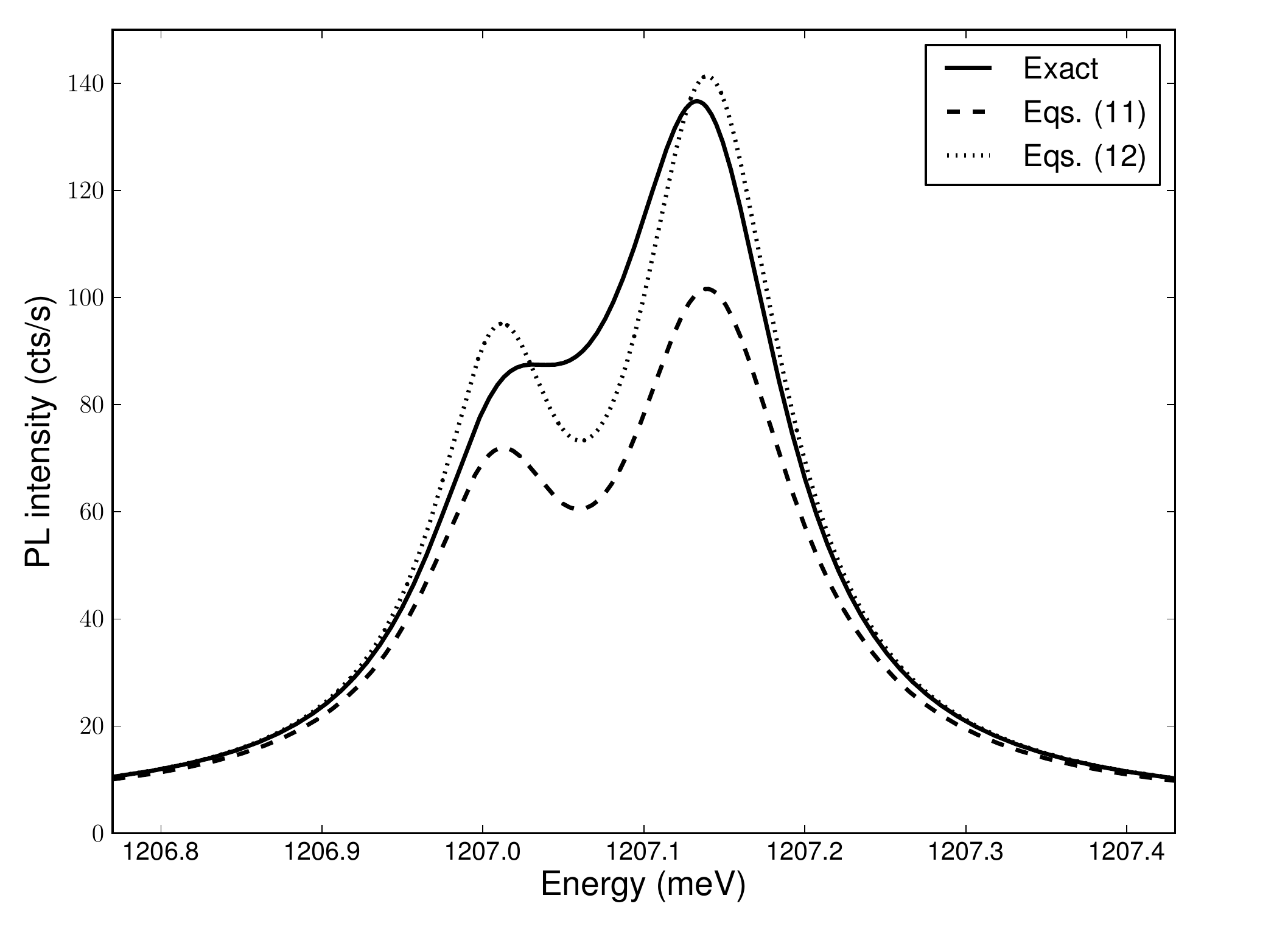} 
\caption{\label{fig:3} Emission spectra calculated with three different methods for the parameters reported by Finley and co-workers \cite{PRL_finley} ($g=59\,\mu$eV/$\hbar$, $P_{a}=4.5\,\mu$eV/$\hbar$, $P_{\sigma}=0.5\,\mu$eV/$\hbar$, $\gamma_{a}=68.0\,\mu$eV/$\hbar$, $\gamma_{\sigma}=0.2\,\mu$eV/$\hbar$, and $\gamma_{\phi}=19.9\,\mu$eV/$\hbar$). The continuous line is the PL spectrum computed with the exact expansion (Eqs. (\ref{QRTeqs})) up to $\Lambda_{25}$. Dashed and doted lines correspond to the spectrum calculated with Eqs. (\ref{app1}) and Eqs. (\ref{app2}), respectively.}
\end{figure}

There are at least two ways of finding an approximation to the emission spectrum. The first one is to approximate the first-order correlation function up to the first excitation manifold\cite{PRA_carmichael}. This leaves us with the following set of differential equations 
\begin{equation}
\begin{split}\label{app1}
 \frac{d}{d\tau}\langle a^{\dagger}(\tau)\rangle&=\left(i\omega_{c}-\Gamma_{1}\right) \langle a^{\dagger}(\tau)\rangle+ig\langle\sigma_{+}(\tau)\rangle\\
 \frac{d}{d\tau}\langle\sigma_{+}(\tau)\rangle&=ig\langle a^{\dagger}(\tau)\rangle+\left(i\omega_{x}-\Gamma_{1}'\right)\langle\sigma_{+}(\tau)\rangle,
\end{split}
\end{equation}
where $\Gamma_{1}=3P_{a}/2+\gamma_{a}/2+P_{\sigma}$ and $\Gamma_{1}'=P_{a}+P_{\sigma}/2+\gamma_{\sigma}/2+\gamma_{\phi}$. Then, the first-order correlation function $G^{(1)}(t,\tau)$ can be found by using the Quantum Regression Theorem with the appropriate initial conditions. The second way is due to Finley  and co-workers\cite{PRL_finley,PRB_finley}, and it combines the first approximation with the exact solution as we show in Appendix \ref{Approxs}. Again, the problem becomes two-dimensional and the spectrum is computed via
\begin{equation}\label{app2}
\begin{split}
  \frac{d}{d\tau}\langle a^{\dagger}(\tau)\rangle&=\left(i\omega_{c}-\Gamma_{2}\right) \langle a^{\dagger}(\tau)\rangle+ig\langle\sigma_{+}(\tau)\rangle\\
  \frac{d}{d\tau}\langle\sigma_{+}(\tau)\rangle&=ig\langle a^{\dagger}(\tau)\rangle+\left(i\omega_{x}-\Gamma_{2}'\right)\langle\sigma_{+}(\tau)\rangle,
\end{split}
\end{equation}
where $\Gamma_{2}=\gamma_{a}/2-P_{a}/2$ and $\Gamma_{2}'=P_{\sigma}/2+\gamma_{\sigma}/2+\gamma_{\phi}$.
Both approximations are good for low pumping rates. However, as displayed in Fig. \ref{fig:3}, they are unable to reproduce the exact spectrum for typical parameters controlling the dynamics of the nC-QD system in real experimental setups.
    
\subsection{Theoretically fitting an experimental emission spectrum}
To validate our model of the polariton bath, we followed a similar procedure to the one developed by Finley and co-workers\cite{PRB_finley, PRL_finley}. By fitting one of their experimental spectra\cite{PRL_finley} we were able to find values for the coupling constant $g$ and the gain and loss rates involved in our master equation (Eq. (\ref{master})). Both the detuning and the cavity mode were experimentally determined by Finley's Group so they are not included in the fitting process and remain fixed at $\Delta=-50\,\mu$eV$/\hbar$ and $\omega_{c}=1207$.$1\,$meV$/\hbar$. We also use their estimated value of the broadening parameter appearing in Eq. (\ref{S(w)}), $\gamma=30\,\mu$eV$/\hbar$\cite{PRL_finley}.\par

Based on the results of the previous section, when fitting an experimental emission spectrum it is of great importance to compute $G^{(1)}(t,\tau)$  using the exact expansion (Eq. (\ref{g1})) since for some range of parameters the two approximations mentioned above are not able to reproduce the actual spectral lines. As a consequence, we do not have an analytical expression for $S(\omega)$ so we cannot use a regular least-squares algorithm to fit the spectrum. To overcome this obstacle we used a Simulated Annealing (SA) algorithm\cite{numerical_recipes} defining the cost function as the sum of square differences between the experimental and theoretical spectral lines
\begin{equation}\label{error}
\mathcal{S}^{2}=\sum_{i}(S(\omega_{i})-S_{exp}(\omega_{i}))^{2}, 
\end{equation}
where $S_{exp}(\omega_{i})$ is the experimental PL intensity at frequency $\omega_{i}$, and the index $i$ runs over the available data. SA has proven to be a fairly good tool to optimize functions in high parametric spaces making it the best choice for our purposes.\par

\begin{figure}
\includegraphics[scale=0.4]{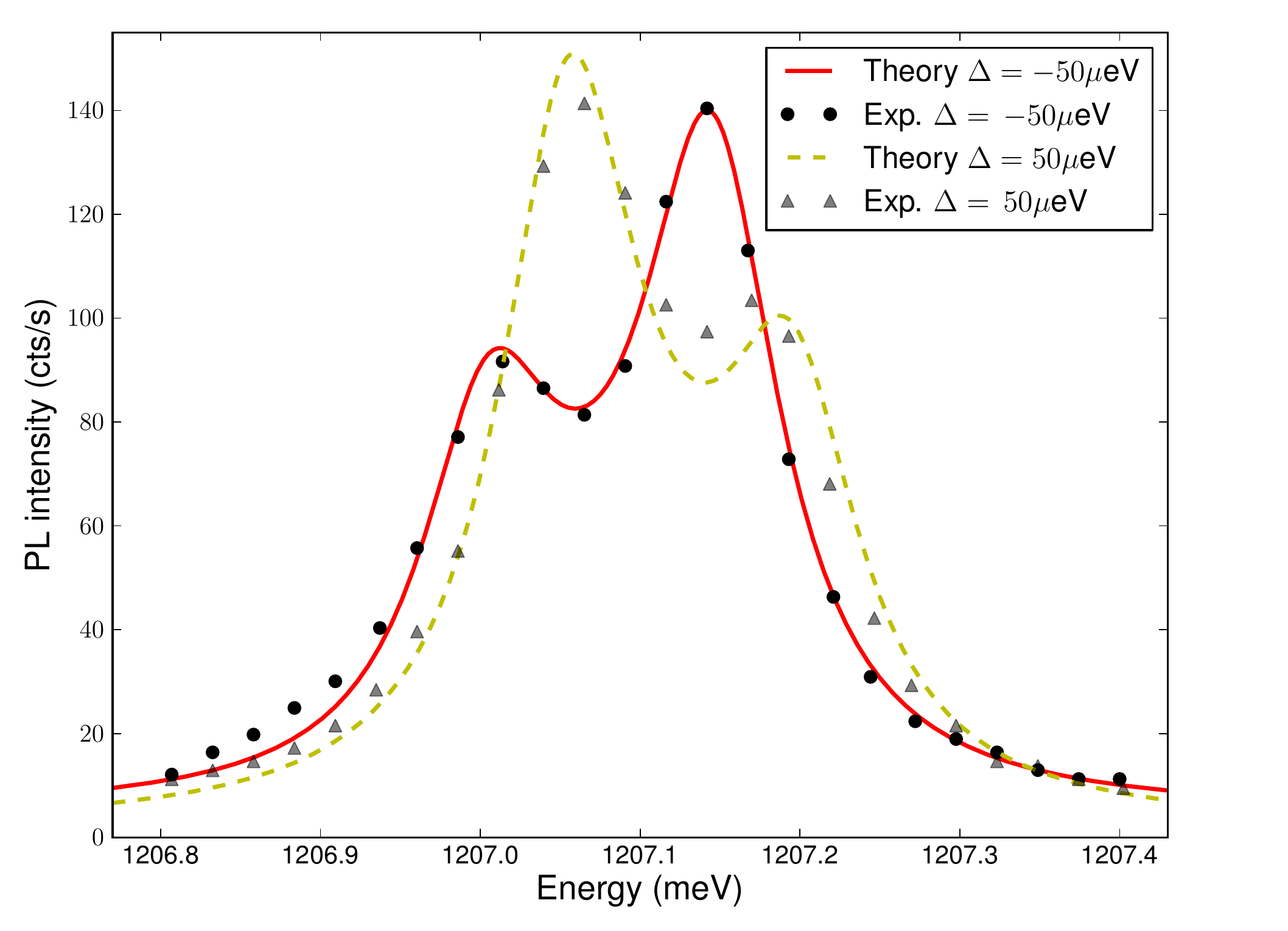} 
\caption{\label{fig:4}(color online) Theoretical fit and control spectra. The continuous red line shows the theoretical PL spectrum  fitting the experimental data (black dots) at $\Delta=-50\,\mu$eV$/\hbar$. The dashed yellow line is the control spectrum and the gray triangles are the experimental intensities at $\Delta=+50\,\mu$eV$/\hbar$. The cost function is $\mathcal{S}=12.6\,$cts/s in the first case, and $\mathcal{S}=29.8\,$cts/s in the second one. The fundamental mode frequency is fixed at $\omega_{c}=1207.1\,$meV$/\hbar$.}
\end{figure}

The best fit (displayed in Fig. \ref{fig:4}) was obtained for $g=67.93\,\mu$eV/$\hbar$, $P_{p}=0.02\,\mu$eV/$\hbar$, $P_{a}=2.27\,\mu$eV/$\hbar$, $P_{\sigma}=2.32\,\mu$eV/$\hbar$, $\gamma_{p}=8.96\,\mu$eV/$\hbar$, $\gamma_{a}=31.82\,\mu$eV/$\hbar$, $\gamma_{\sigma}=10.12\,\mu$eV/$\hbar$, and $\gamma_{\phi}=7.40\,\mu$eV/$\hbar$. These values show that polariton pumping and loss rates are comparable to the other parameters, indicating that the polariton bath we proposed has a significant role in the dynamics of the nC-QD system. A conclusion that can be drawn given the quality of our fit.\par

Additionally, to strengthen our result we use a control experimental spectrum from Ref. \onlinecite{PRL_finley} with $\Delta =+50\,\mu$eV$/\hbar$. The experimental data along the theoretical curve computed with the same fitting parameters, are also illustrated in Fig. \ref{fig:4}. This control spectrum shows that the parameters obtained are also good to describe the characteristic asymmetric double peak features near resonance. Note that the small blue shift in the spectral lines is consistent with the eigenenergies of the ``dressed'' states.\par

\section{Conclusions}\label{sec:conclusions}
To summarize, we have presented a phenomenological model for a polariton bath interacting with a nC-QD system. By theoretically fitting a photoluminescence spectrum, we were able to show that our model agrees with the experiment for non-zero values of the polariton pumping and loss rates. These values have the same order-of-magnitude of the rest master equation parameters, showing that at least from an experimental point of view the polariton gain and loss processes are feasible in real nC-QED systems. Additionally, we have shown the importance of computing the PL spectrum with the exact dynamics of the first-order correlation function. The two approximations we have discussed, fail to give an accurate description of the nC-QD system in real experimental conditions.

\section*{Acknowledgments} We gratefully  acknowledge partial financial support from Direcci\'on de Investigaci\'on - Sede Bogot\'a, Universidad Nacional de Colombia (DIB-UNAL), and CODI at Universidad de Antioquia. The authors thank Nicol\'as Quesada, Carlos A. Vera, Dr. Juliana Restrepo, Prof. Paulo S. S. Guimar{\~a}es and Prof. Karen M. Fonseca Romero for useful discussions and critical reading of the manuscript. B.A.R acknowledge the support of Departamento de F{\'i}sica, Universidade Federal de Minas Gerais for a short term visit.

\appendix

\section{Density operator matrix elements equations}\label{Matrix}
Denoting matrix elements by $\rho_{Im,Jn}=\langle Im|\rho|Jn\rangle$, with $I,J\in\{G,X\}$, and using Eq. (5), is straightforward to show that   

\begin{widetext}

\begin{subequations}\label{matrix}
 \begin{align}
  \frac{d}{dt}\rho_{Gn,Gn}=\ &n\,(P_{a}+P_{p})\,\rho_{Gn-1,Gn-1}+ nP_{p}\,\rho_{Xn-2,Xn-2}\notag\\
  -& \bigl\{(n+1)P_{a}+n\gamma_{a}+P_{\sigma}+2(n+1)P_{p} + (2-\delta_{n1})n\,\gamma_{p}\bigr\}\rho_{Gn,Gn} + ig \sqrt{n} \,\rho_{Gn,Xn-1}\notag\\
   -& ig\sqrt{n}\,\rho_{Xn-1,Gn} + (n+1)(\gamma_{a}+\gamma_{p})\,\rho_{Gn+1,Gn+1} + \{\gamma_{\sigma}+(n+1)\gamma_{p}\}\,\rho_{Xn,Xn}\,,\label{rGN}
\end{align}
\end{subequations}
\begin{subequations}
\begin{align}
  \frac{d}{dt}\rho_{Xn,Xn}=\ & \{P_{\sigma}+(n+1)P_{p}\}\,\rho_{Gn,Gn} + \{nP_{a}+(n+1)P_{p}\}\,\rho_{Xn-1,Xn-1}\notag\\[4pt]
  -&\bigl\{\!(n+1)P_{a}+n\gamma_{a}+\gamma_{\sigma}+2(n+2)P_{p} + (2-\delta_{n0})(n+1)\gamma_{p}\!\bigr\}\rho_{Xn,Xn}
  - ig\sqrt{n+1}\,\rho_{Gn+1,Xn}\notag\\
  +& ig\sqrt{n+1}\,\rho_{Xn,Gn+1} + (n+2)\gamma_{p}\,\rho_{Gn+2,Gn+2} + \{(n+1)\gamma_{a}+(n+2)\gamma_{p}\}\,\rho_{Xn+1,Xn+1}\label{rXn}\,,\tag{\ref{matrix}b}
\end{align}
\end{subequations}
\begin{subequations}
\begin{align}
  \frac{d}{dt}\rho_{Gn+1,Xn}=\ & \sqrt{n(n+1)}\,P_{a}\,\rho_{Gn,Xn-1} + \sqrt{(n+1)(n+2)}\,\gamma_{a}\,\rho_{Gn+2,Xn+1}\notag\\
  +&\biggl\{i\Delta-(2n+3)\frac{P_{a}}{2}-(2n+1)\frac{\gamma_{a}}{2}-\frac{P_{\sigma}}{2}-\frac{\gamma_{\sigma}}{2}-\gamma_{\phi}
  -2(n+2)P_{p} - (2-\delta_{n0})(n+1)\gamma_{p}\biggr\}\,\rho_{Gn+1,Xn}\notag\\
  +& ig\sqrt{n+1}\,\rho_{Gn+1,Gn+1} - ig\sqrt{n+1}\,\rho_{Xn,Xn}\label{rGn+1Xn}\,,\tag{\ref{matrix}c}
 \end{align}
\end{subequations}

\noindent
Since $\rho_{Xn,Gn+1}=\rho_{Gn+1,Xn}^{\ast}$, the equation governing time evolution for $\rho_{Xn,Gn+1}$ is found by taking the complex conjugate of Eq. (\ref{matrix}c). As mentioned above, this linear set of differential equations is quite useful to find the time evolution of the density operator $\rho$.

\section{QRT equations}\label{QRT}
Expectation values for $a^{\dagger}_{Gn}$, $a^{\dagger}_{Xn}$, $\sigma^{\dagger}_{n}$, $\varsigma_{n}$ can be written in terms of density operator matrix elements using the master equation (Eq. \ref{master}). Their time evolution can be easily found to be
\begin{small}
\begin{subequations}\label{QRTeqs}
\begin{align}
 \frac{d}{d\tau}\langle a_{Gn}^{\dagger}(\tau)\rangle=&\biggl\{i\omega_{c}-\frac{P_{a}}{2}(2n+3)-\frac{\gamma_{a}}{2}(2n+1)
 -P_{\sigma}-P_{p}(2n+3)-\frac{\gamma_{p}}{2}\bigl[(2-\delta_{n1})n+(2-\delta_{n0})(n+1)\bigr]\biggr\}\langle a_{Gn}^{\dagger}(\tau)\rangle\\
 +&P_{a}\sqrt{n(n+1)}\,\langle a_{Gn-1}^{\dagger}(\tau)\rangle-ig\sqrt{n}\,\langle\varsigma_{n}(\tau)\rangle
 +\gamma_{a}\sqrt{(n+1)(n+2)}\,\langle a_{Gn+1}^{\dagger}(\tau)\rangle
 +\gamma_{\sigma}\langle a_{Xn}^{\dagger}(\tau)\rangle+ig\sqrt{n+1}\,\langle\sigma^{\dagger}_{n}(\tau)\rangle;\notag
\end{align}
\end{subequations}
\begin{subequations}
\begin{align}
\frac{d}{d\tau}\langle a_{Xn}^{\dagger}(\tau)\rangle&=\biggl\{i\omega_{c}-\frac{P_{a}}{2}(2n+3)-\frac{\gamma_{a}}{2}(2n+1)
-\gamma_{\sigma}-P_{p}(2n+5)-\frac{\gamma_{p}}{2}\bigl[(2-\delta_{n0})(n+1)+2(n+2)\bigr]\biggr\}\langle a_{Xn}^{\dagger}(\tau)\rangle \tag{\ref{QRTeqs}b}\\
+P_{a}&\sqrt{n(n+1)}\,\langle a_{Xn-1}^{\dagger}(\tau)\rangle+ig\sqrt{n+2} \,\langle\varsigma_{n+1}(\tau)\rangle+\gamma_{a}\sqrt{(n+1)(n+2)}\,\langle a_{Xn+1}^{\dagger}(\tau)\rangle
+P_{\sigma}\langle a_{Gn}^{\dagger}(\tau)\rangle-ig\sqrt{n+1}\,\langle\sigma^{\dagger}_{n+1}(\tau)\rangle;\notag
\end{align}
\end{subequations}
\begin{subequations}
\begin{align}
\frac{d}{d\tau}\langle\sigma_{n}^{\dagger}(\tau)\rangle&=\biggl\{i\omega_{x}-P_{a}(n+1)-\gamma_{a}n
-\frac{P_{\sigma}}{2}-\frac{\gamma_{\sigma}}{2}-\gamma_{\phi}-P_{p}(2n+3)
-\frac{\gamma_{p}}{2}\bigl[(2-\delta_{n0})(n+1)+(2-\delta_{n1})n\bigr]\biggr\}\langle\sigma_{n}^{\dagger}(\tau)\rangle\notag\\
&+P_{a}n\langle\sigma_{n-1}^{\dagger}(\tau)\rangle+\gamma_{a}(n+1)\langle\sigma_{n+1}^{\dagger}(\tau)\rangle
+ig\sqrt{n+1}\,\langle a_{Gn}^{\dagger}(\tau)\rangle-ig\sqrt{n}\,\langle a_{Xn-1}^{\dagger}(\tau)\rangle; \tag{\ref{QRTeqs}c}
\end{align}
\end{subequations}
\begin{subequations}
\begin{align}
\frac{d}{d\tau}\langle\varsigma_{n}(\tau)\rangle&=\biggl\{i(2\omega_{c}-\omega_{x})-P_{a}(n+1)-\gamma_{a}n
-\frac{P_{\sigma}}{2}-\frac{\gamma_{\sigma}}{2}-\gamma_{\phi}-P_{p}(2n+3)
-\frac{\gamma_{p}}{2}\bigl[(2-\delta_{n0})(n+1)+(2-\delta_{n1})n\bigr]\biggr\}\langle\varsigma_{n}(\tau)\rangle\notag\\
&+P_{a}\sqrt{n^{2}-1}\,\langle\varsigma_{n-1}(\tau)\rangle+\gamma_{a}\sqrt{n(n+2)}\,\langle\varsigma_{n+1}(\tau)\rangle
+ig\sqrt{n+1}\,\langle a_{Xn-1}^{\dagger}(\tau)\rangle-ig\sqrt{n}\,\langle a_{Gn}^{\dagger}(\tau)\rangle. \tag{\ref{QRTeqs}d} 
\end{align}
\end{subequations}
\end{small}
\end{widetext}
\noindent
From Eqs. (\ref{QRTeqs}), it becomes clear that the position of spectral peaks solely depends on $g$, $\omega_{x}$, $\omega_{c}$. This means that dissipation and pumping rates only modify their intensity and width. This is expected since population dynamics and decoherence properties are ruled by these rates.\\\par

\section{Approximations to the emission spectrum: Derivations}\label{Approxs}
Both approximations to the emission spectrum discussed in section \ref{sec_apps} can be derived from Eqs. (\ref{QRTeqs}), which describe the exact dynamics of the first-order correlation function $g^{(1)}(t,\tau)$. The first approximation is straight forward. Up to $\Lambda_{1}$, we have $\langle a^{\dagger}\rangle\approx \langle a^{\dagger}_{G0}\rangle$ and $\langle\sigma_{+}\rangle\approx\langle\sigma^{\dagger}_{0}\rangle$ so that other expectation values in Eqs. (\ref{QRTeqs}) become zero, leaving the two-dimensional system in Eqs. (\ref{app1}). To derive the second approximation we need to find the exact time derivatives of $\langle a^{\dagger}\rangle$ and $\langle\sigma_{+}\rangle$. From Eqs. (\ref{QRTeqs}), summing over all values of $n$ and simplifying, we find
\begin{equation}\label{dadt}
\begin{split}
 \frac{d}{d\tau}\langle a^{\dagger}(\tau)\rangle&=\left\{i\omega_{c}+\frac{P_{a}}{2}-\frac{\gamma_{a}}{2}\right\}\times\\
 &\sum_{n=0}^{\infty}\sqrt{n+1}\,\bigl(\langle a_{Gn}^{\dagger}(\tau)\rangle+\langle a_{Xn}^{\dagger}(\tau)\rangle\bigr)\\
 &+ig\sum_{n=0}^{\infty}\langle\sigma^{\dagger}_{n}(\tau)\rangle,
\end{split}
\end{equation}\vspace*{-22pt}
\begin{equation}\label{dsdt} 
\begin{split}
\frac{d}{d\tau}\langle\sigma_{+}(\tau)\rangle&=\biggl\{i\omega_{x}-\frac{P_{\sigma}}{2}-\frac{\gamma_{\sigma}}{2}-\gamma_{\phi}\biggr\}\sum_{n=0}^{\infty}\langle\sigma_{n}^{\dagger}(\tau)\rangle\\
 -ig&\sum_{n=0}^{\infty}\sqrt{n+1}\,\bigl(\langle a_{Xn}^{\dagger}(\tau)\rangle-\langle a_{Gn}^{\dagger}(\tau)\rangle\bigr),
\end{split}
\end{equation}
It is worth noting that this pair of equations are exact and cannot be found by doing the first excitation manifold approximation. From Eq. (\ref{a}) and since $\sigma_{+}=\sum_{n}\sigma_{n}^{\dagger}$, we have that Eq. (\ref{dadt}) is precisely the first line in Eqs. (\ref{app2}). However, to obtain the second line, it is necessary to approximate the second sum in Eq. (\ref{dsdt}) to $\Lambda_{1}$, i.e.,   
\begin{equation}\label{sumapp}
\begin{split}
\sum_{n=0}^{\infty}\sqrt{n+1}\,\bigl(\langle a_{Xn}^{\dagger}(\tau)\rangle-\langle a_{Gn}^{\dagger}(\tau)\rangle\bigr)&\approx-\langle a^{\dagger}_{G0}\rangle,\\ 
&\approx-\langle a^{\dagger}\rangle.
\end{split}
\end{equation}
This may be objectionable, since to find Eq. (\ref{dsdt}) we had to sum over all excitation manifolds and now we are approximating the last sum to $\Lambda_1$.

\bibliography{biblio_paper}
\end{document}